\documentclass[11pt, a4paper]{article} 
\usepackage[dvips]{graphicx}
\usepackage{epsfig}
\usepackage{palatino}
\usepackage{amssymb}
\usepackage{amsmath}
\usepackage{setspace}
\usepackage[displaymath]{lineno}
\date{}

\begin{document}

\title{A model for the time uncertainty measurements in the Auger 
surface detector array}
\author
       {C. Bonifazi$^{1,}$\footnote{bonifazi@cbpf.br, corresponding author.}~, A. Letessier-Selvon$^{1,2,}$\footnote{Antoine.Letessier-Selvon@in2p3.fr}~, and E.M. Santos$^{1,}$\footnote{emoura@cbpf.br} \\ for the Pierre Auger Collaboration}  
\maketitle       
{\small \noindent
$^1$~Centro Brasileiro de Pesquisas F\'isicas, Rua Dr. Xavier Sigaud
150, Rio de Janeiro, 22290-10, Brazil, \\
$^2$~Laboratoire de Physique Nucl\'eaire et des Hautes \'Energies, T33
RdC, 4 place Jussieu, 75252 Paris Cedex 05, France\\
} 

\begin{abstract}
The precise determination of the arrival direction of cosmic rays is a 
fundamental prerequisite for the search for sources or the study
of their anisotropies on the sky. One of the most important aspects 
to achieve an optimal measurement of these directions is to properly
take into account the measurement uncertainties in the estimation
procedure. In this article we present a model for the uncertainties 
associated with the time measurements in the Auger surface detector
array. We show that this model represents well the measurement
uncertainties and therefore provides the basis for an optimal 
determination of the arrival direction. With this model and a
description of the shower front geometry it is possible to 
estimate, on an event by event basis, the uncertainty associated with 
the determination of the arrival directions of the cosmic rays.

\end{abstract}


\vspace{-0.6cm}

\section{Introduction}

The nature and origin of ultra high energy cosmic rays (UHECR) is a
continuing puzzle of modern astrophysics~\cite{review}. Essential 
to the study of their origin is the measurement of their arrival
directions with a careful evaluation of their measurements uncertainties. 

The Pierre Auger Observatory~\cite{NIM} is designed to measure the 
flux, arrival direction distribution, and composition of the
UHECR. In order to have full sky coverage, it will have two
components, one in each Hemisphere. The Southern site of the
Observatory is in its last stages 
of construction in Malarg\"ue, Argentina, and has been acquiring 
data since 2004. When completed, it will be composed of 1,600 water
Cherenkov detectors (WCD) spread over 3,000~km$^2$ with the atmosphere
above viewed by 24 fluorescence telescopes placed in 4 buildings.

Each WCD consists of a rotomolded polyethylene tank with an area of 
10~m$^2$ and 1.55~m high, enclosing a
Tyvek\textsuperscript{\textregistered}~liner filled with 12,000~l of 
high purity water. The Cherenkov light is detected by three large
9"~photomultipliers (PMTs) viewing the water tank from the top. Each
detector is an autonomous unit. The signals from the PMTs are
processed by the local electronics. The power of the whole system is
provided by two batteries connected to two solar panels. The time
synchronization is done by commercial GPS receivers. There is
also a communication system between each detector and the Central
Station of the Observatory.  

The UHECRs are detected through the particle cascade they produce in 
the atmosphere. With the Auger surface detector (SD) we determine
their arrival direction from a fit to the particle arrival times on the 
ground.

During the development in the atmosphere, the air shower increases in 
size up to a maximum. The bulk of the particles reaching the ground are 
grouped in a thick (hundreds of nano-seconds to a few micro-seconds,
e.g. 100~m to 1000~m) nearly spherical pancake. The thickness depends on the 
distance to the shower axis, while the radius of curvature of the
sphere that can be used to describe the shower front approximately depends
on the altitude of the shower maximum.   


The precision in the cosmic ray arrival direction achieved from the SD
reconstruction depends on the clock precision of the detector and on 
the fluctuations in the first particle arrival time at
ground. These fluctuations increase with the shower thickness and
decrease with the particle density. Therefore, the shower front time
will be measured with less precision for a detector located
further away from the core, where particle densities are small and the
shower thickness is large, than for the same detector closer to the core 
of the same shower.

To estimate the primary direction of the primary cosmic ray with the 
maximum precision, one must, among other things, adequately model the
measurement uncertainties of each individual tank participating in the
event. These measurement uncertainties depends on the shower
properties at each tank location. With an adequate model for the
individual measurements, one is able to weight the
contribution in the determination of the arrival direction correctly.  

The model of the shower front used in the minimization procedure, be
it spherical, parabolic, or even planar also influences the
uncertainty in the arrival direction determination, but not as 
much as the time measurement precision. We will show in
section~\ref{prob} that a parabolic model for the shower front
adequately describes the data.

In section~\ref{model} we describe a model for the shower front
particle distribution from which we derive the uncertainty in the  
time measurement at each tank. In section~\ref{m-imp} we implement this
model for the Auger WCDs. In section~\ref{data} we compare it to 
measurements obtained using two pairs of adjacent detector stations 
located in the Auger surface array. Finally, in section~\ref{prob} we 
further validate our model by studying the $\chi^2$ probability
distribution of the geometrical reconstruction of Auger events.

\section{A model for the measurement uncertainties}\label{model}

As an estimator for the shower front arrival time ($T_s$) at each
detector location on the ground it is customary to use the time of 
the first particle entering the tank. As discussed in the 
introduction, assuming one has a reasonable description of the shower
front geometry, it is the estimation accuracy on $T_s$ at each tank 
that will determine the overall precision with which the Auger SD 
can reconstruct the primary cosmic ray arrival direction. 

At a given point on the ground, we assume that the particle distribution
in the shower front can be described as a Poisson process with
parameter ${\mathcal \tau}$ starting at time $t_0$. Of course, both
$\mathcal \tau$ and $t_0$ are functions of the shower characteristics 
at the considered ground location. In reality, the particle
distribution in the shower front is not a uniform Poisson process. At
the beginning of the front the particle arrival frequency is larger
than towards the end. It is however sufficient to assume that the
frequency is constant over some interval large enough to have a good
fraction of the total number of particles reaching the ground but less
than the total thickness of the shower front at that location. 

The first particle arrival time as recorded by the detector clock is
then given by:
\begin{equation}
T_s = t_0 + T_1
\end{equation}
where $T_1$ follows an exponential distribution function with decay
parameter ${\mathcal \tau}$. 

\begin{figure}
  \begin{center}
    \includegraphics*[width=0.7\textwidth]{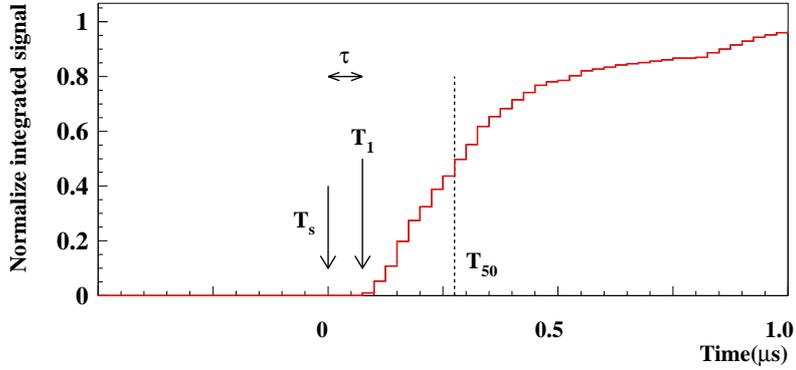}
    \vspace{-0.7cm}
    \caption {Example of an integrated FADC trace for a detector at 
      ~700~m from the shower core position. $T_s$ is the shower front
      arrival time, $T_1$ is the arrival time of the first particles, 
      and $T_{50}$ is the time it takes to reach 50\% of the total
      integrated signal in the trace. 
      \label{fig:times}}
  \end{center}
\end{figure}

Note that, by construction $T_s$ (our shower front arrival time
estimator) is always larger than the true (unknown) shower front time
$t_0$. $T_s$ is a biased estimator of $t_0$ with a bias given by the
expectation of $T_1$, $E[T_1] = \mathcal \tau$.  If $\mathcal \tau$ is
known it would be interesting to construct a new shower front
estimator as $T'_s = T_s - {\mathcal \tau}$. However, ${\mathcal
  \tau}$ can only be estimated from the shower parameters measured at
the ground and is therefore subject to fluctuations. The new estimator
$T'_s$ has therefore slightly larger fluctuations than
$T_s$~\footnote{When the radius of curvature of the shower front is
  estimated during the geometrical minimization (which happens for tank
multiplicities larger than 5) the bias is absorbed in the radius of 
curvature. For low multiplicity events with only 3 tanks where there
is no constraint to determine the geometry, this bias induces a
systematic in the pointing but much smaller than the pointing
resolution itself.}. For an schematic view of our notation see
figure~\ref{fig:times}. 
     
The variance of $T_s$ ($V[T_s]$) is given by the sum of the detector 
clock precision ($b^2$) and of the variance of the random variable
$T_1$, $V[T_1] = \mathcal \tau^2$. In practice $\mathcal \tau$ is
unknown and must be estimated from the data itself. We reverted to use 
(see section~\ref{m-imp}) as an estimator of $\mathcal \tau$ the ratio of
the time of arrival of the $k$-th particle ($T_k$) to $k$: $\hat{\tau} =
T_k/k$. The variance of $T_s$ 
then becomes\footnote{$V[T_1]$ is not strictly $\hat{\tau}^2 =
(T_k/k)^2$ but $\hat{\tau}^2(k-1)/(k+1)$ because $T_k$ is measured
from the data themselves.}: 
\begin{equation}
  V[T_s] = \left( \frac {T_k}{k}\right )^2~\frac {k-1}{k+1} + b^2,
  \label{eq:var0}
\end{equation}
where $b^2$ represents the GPS uncertainty and the resolution of 
the flash analog-to-digital converters (FADCs) of the tanks.

\section{Model implementation}\label{m-imp}

To estimate equation~\ref{eq:var0} for each tank participating in the 
event we want to reconstruct, we must define ${\mathcal T_k}$ and $k$ 
from the data measured at the tank. For $k$ we choose to use the total
number of particles entering the tank ($n$ in the following).

The shower signal is recorded in the Auger WCD by the FADCs, which
give the signal values in bins of 25~ns as a function of time. From
these FADC traces one can build several time parameters such as the 
time it takes to reach 50\% of the total integrated signal in the
trace: $T_{50}$ (see figure~\ref{fig:times}). Following our remark in
section~\ref{model}, we decided to use $2T_{50}$ for $T_n$ rather than
$T_{100}$ to account for the fact that the Poisson process is not
uniform over the whole shower front. 

To calculate the total number of particles, $n$, we assume that all
particles hit the detector with the same direction as the shower
axis, and that the muons (which arrive first) are the ones that 
contribute most to the time measurements~\cite{ls}. Then, we 
obtain $n$ as the ratio 
between the total integrated signal in the FADC traces in units of 
vertical equivalent muons~\cite{calib-paper} (the signal deposited by 
a central and vertical downgoing muon in a tank), divided by the average 
muon track length $TL(\theta)$ of a shower with zenith angle $\theta$,
normalize to the vertical height ($h$):
\begin{equation}
n = \frac{S_{VEM}}{TL(\theta)/h}.
\label{eq:n}
\end{equation}

The average track length $TL(\theta)$ can be computed as the ratio of 
the detector volume ($V$) and the area ($A$) subtended by the arriving 
particles at zenith angle $\theta$. For the Auger cylindrical 
tanks, it is given by: 
\begin{equation}
  TL(\theta) = \frac{V}{A} = \frac{\pi r h}
  {\pi r \cos(\theta) + 2 h \sin(\theta)},
\label{eq:fittl}
\end{equation}
\noindent
where $r$ = 1.8~m is the detector radius, and $h$ = 1.2~m is the water 
height in the detector.

We further introduce a scale parameter $a$ to account for the various 
approximations made to build $V[T_s]$ and finally define it as:
\begin{equation}
  V[T_s] = a^2~\left(\frac {2~T_{50}}{n}\right)^2~\frac {n-1}{n+1} + b^2.
  \label{eq:var}
\end{equation}

Parameters $a$ and $b$ can be determined from the data of the two pairs of 
adjacent stations available in the Auger array (see section
\ref{doublets}). We expect, however, $a$ to be close to unity and to
$b$ be given by our GPS system resolution \cite{GPS} (10~ns) and the 
resolution of our FADC traces (25/$\sqrt{12}$~ns), giving~$b\simeq 12$~ns.

\section{Comparison with data}\label{data}
\subsection{Fitting the doublet data}\label{doublets}

Two pairs of adjacent surface detector stations separated by 11~m
(``doublets'') have been installed in the field of the Auger
Observatory. These pairs enable comparison of timing and signal 
accuracy measurements.

To adjust the constants $a$ and $b$ we used all the T4~\cite{trigger}
events from April 2004 to the end of December 2006, with at least one 
doublet information available. We defined the time difference 
$\Delta T = dT_1 - dT_2$ where $dT_1$ ($dT_2$) is the time 
residual of the first (second) twin tank to the fitted shower 
front. Note that using the residual difference assures that $\Delta T$ 
does not depend on the global shower front shape, since the twin tanks 
are only 11~m apart. 

The knowledge of $\Delta T$ only requires the knowledge
of $\theta$ and $\phi$ with moderate precision (a few degrees is
enough). We further required $\mid\Delta T\mid$ to be smaller than 
200~ns to avoid the tails of the distribution and also a good 
overall reconstruction. With these cuts we had 2037 events to fit 
our two parameters $a$ and $b$.

\begin{figure}[t] 
  \vspace{-1.2cm}
  \begin{center}
    \includegraphics*[width=0.75\textwidth]{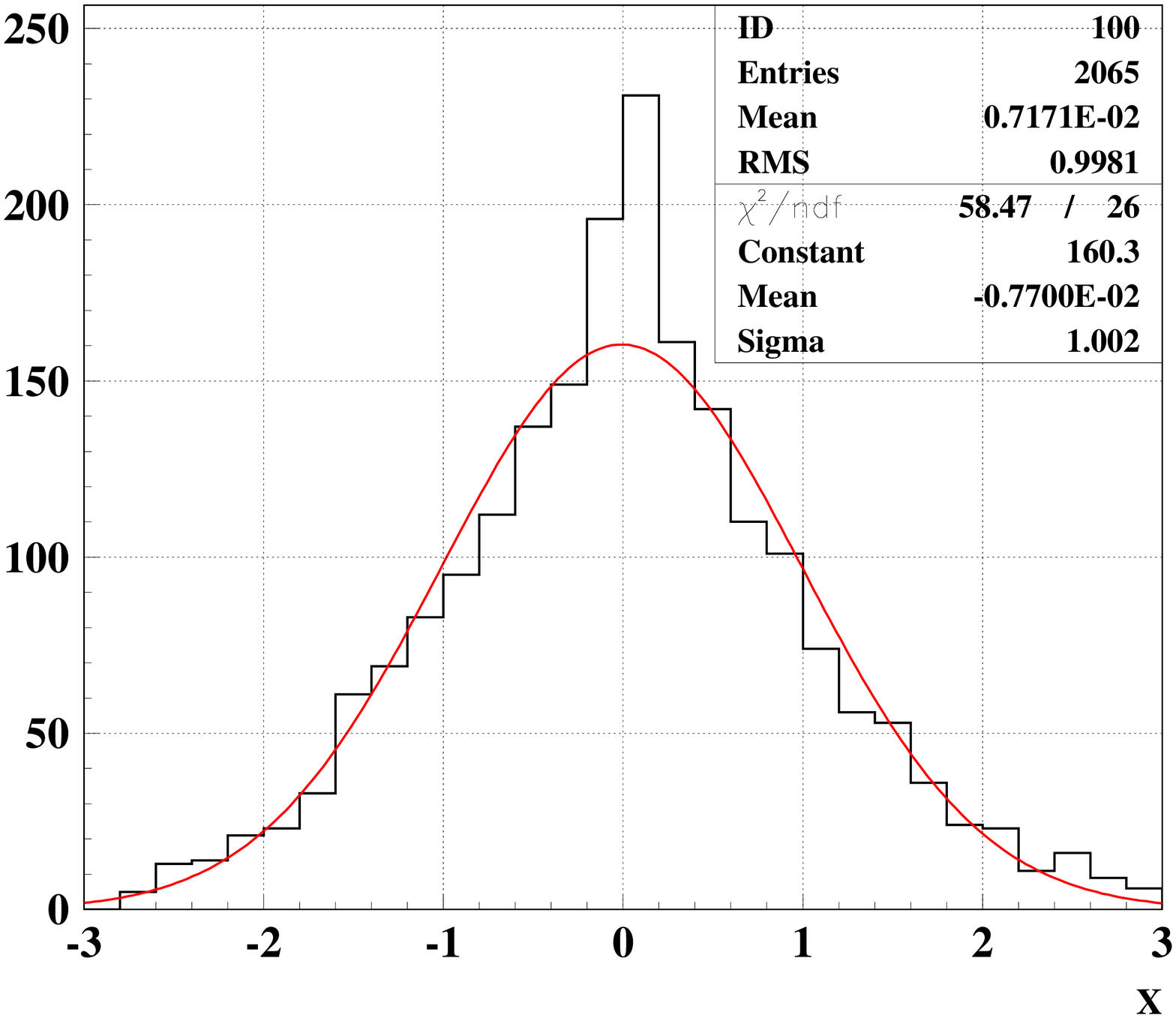}
    \caption{\small {Distribution of $\Delta T/\sqrt(V[\Delta
	  T])$ where $V[\Delta T] = V[T_1] + V[T_2]$ with $T_1$ ($T_2$)
	calculated using equation~\ref{eq:var} for the first (second) twin
	tank, with $a = 1$ and $b = 12$~ns.} \label{fig:dtvar}}
  \end{center}
\end{figure}

Since the distribution of $\Delta T/\sqrt{V[\Delta T]}$ 
is close to a Gaussian (see figure~\ref{fig:dtvar}), we fit the 
parameters by maximizing the following likelihood function 
($\emph L$):
\begin{equation}
\emph{L} =  \prod_{k=1}^N \frac{1}{\sqrt{2\pi V[\Delta T_k]~}}~~
e^{-\frac{\Delta T^2_k}{2~V[\Delta T_k]}}
\label{eq:likelihood}
\end{equation}
\noindent
where $N$ is the total number of events and $V[\Delta T_k] =
V[T_{1,k}]+V[T_{2,k}]$ is calculated from the variance of  $T_s$ for
 each member of the doublet and for each event $k$ using equation
\ref{eq:var}. Maximizing $\emph L$ is equivalent to minimizing
$\emph Z = - 2 \ln(\emph L)$, which was done using the {\em
Minuit} package. We obtained:
\[
\begin{array}{ccrcl}
  a^2 &=& 0.97 \pm 0.04,~~~~~
  \\
  b^2 &=& \left( 132 \pm 11 \right) ~\mbox{\rm ns}^2.
\end{array}
\]
These results are in remarkable agreement with the values expected from 
the model definition: $a^2~=~1$ and $b^2~=~144~ns^2$.

As mentioned earlier, we expect the distribution of the variable
$\Delta T$ to be essentially insensitive to the details of the
geometrical reconstruction used to derive it. This is because the
two tanks in a doublet are very close together and $\Delta T$ can be
computed with moderate precision based on the shower angles (a few
degrees) and regardless of the general structure of the shower
front. We verified that indeed the parameters $a$ and $b$ did not
depend on the nature and quality of the initial geometrical fit used
when analyzing the doublet data. We used, for example, values for $a$
and $b$ different from 1 and 12 during the shower geometrical fit, and
also replaced our time variance model by a constant, we always
obtained the same final parameters values (within the
uncertainties). This shows that using the doublets and the estimator
$\Delta T$ we are really independent of the details of the geometrical
reconstruction used to determined the model parameters. 

\subsection{Model quality}

The value of $\emph L $ at the minimum (10.9 per degree of freedom)
is quite meaningless (unlike a $\chi^2$  value), therefore we need to
evaluate the quality of the fit using alternative tests.

We computed the quality $\chi^2 =\sum_{k=1}^{ndof}\frac{\Delta
T_k^2}{V[\Delta T_k]}$, which should be close to the number of degrees
of freedom if $V[\Delta T_k]$ properly models the variance of $\Delta
T$. We obtained at the minimum $\chi^2 / ndof =$~1.00097. 


We also generated a Monte Carlo sample based on the doublet data. For
each event with a doublet we used the values of $T_{50}$ and signal
($S$) of the doublet members and draw a new time value according to
our shower front model, plus a 25~ns uniform distribution
corresponding to our FADC bin size and a Gaussian distribution with
$\sigma = 10$~ns corresponding to the GPS clock accuracy. We then
calculated $\Delta T$ as the difference of these two times and
reproduced our analysis on this MC sample. 

From the minimization of $\emph Z$, we obtain on the MC sample:
\[
\begin{array}{ccr}
  a^2&=&0.74 \pm 0.05,~~~~~
  \\
  b^2&=&\left(215 \pm 28\right)~\mbox{\text ns}^2.
\end{array} 
\]
with a value of $\emph Z = 10.6$ and $\chi^2 = 1.0011$ at the
minimum, which are similar to what we obtained with the real data
and also corresponds to the injected parameters. 
    
\begin{figure}[t]
  \begin{center}
    \includegraphics*[width=0.8\textwidth]{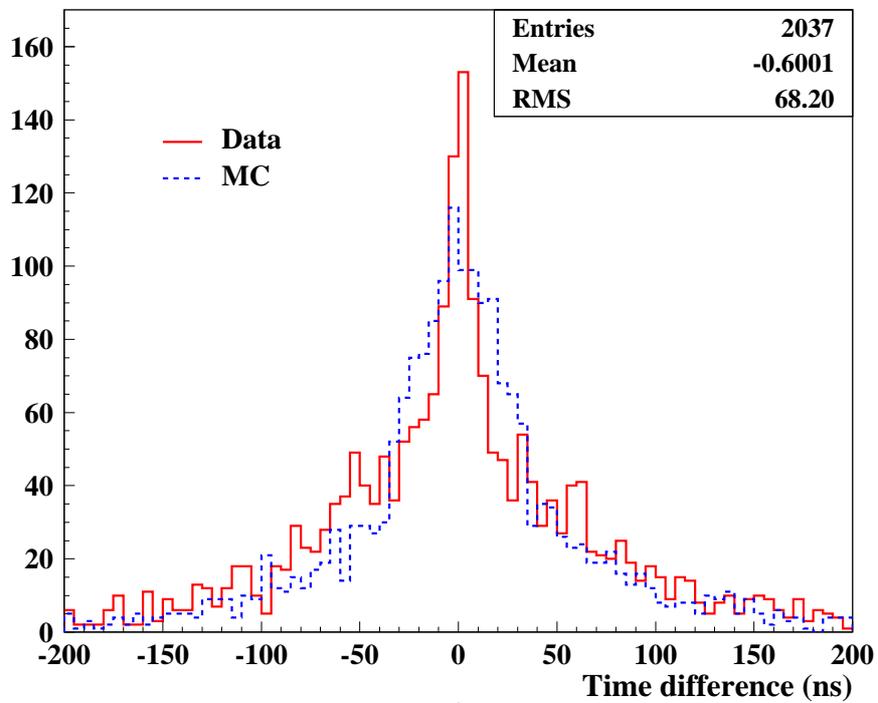}
    \caption{\small {Comparison of the $\Delta T$ distributions for
	doublet data (red-solid) and for MC events (blue-dashed).}
      \label{fig:compara}}
  \end{center}
\end{figure}

In addition, in figure \ref{fig:compara} we plot for comparison 
$\Delta T$ distributions for the doublet data (red-solid) and 
for the Monte Carlo (blue-dashed). As can be seen, the agreement is good.

\section{Validation}\label{prob}

If the time variance model describes correctly the measurement
uncertainties, the distribution of $\Delta T/\sqrt{V[\Delta T]}$, where
$V[\Delta T]~=~V[T_1^{(1)}]~+~V[T_1^{(2)}]$, should have unit variance.

\begin{figure}[t]
  \vspace{-2.5cm}
      \begin{center}
	\includegraphics*[width=0.7\textwidth]{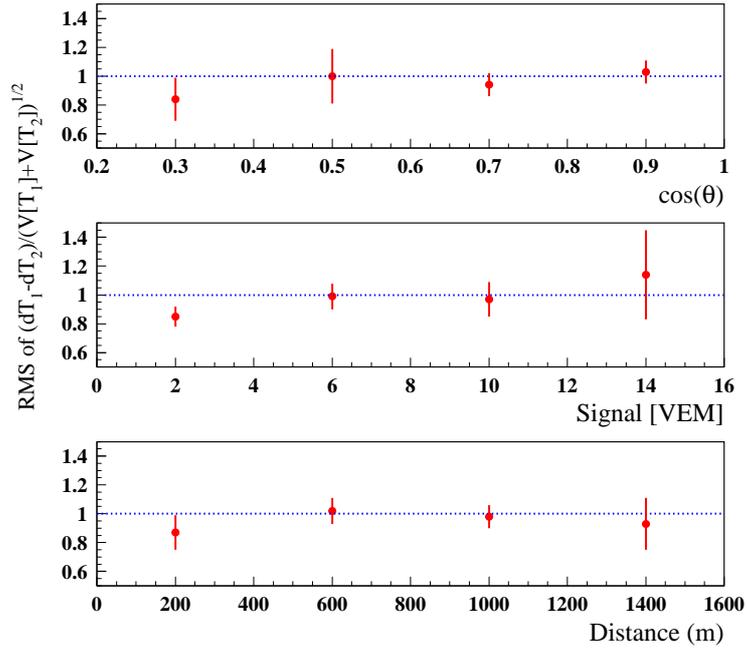}
	\caption {The RMS of the distribution of $\Delta T/\sqrt{V[\Delta T]}$,
          as a function of the shower zenith angle (top), the average
          signal in the doublet detectors (middle), and the distance to
          the shower core (bottom). \label{fig:doublets}}
      \end{center}
\end{figure}

We show on figure \ref{fig:doublets} the RMS of the distributions of
$\Delta T/\sqrt{V[\Delta T]}$ for various bins in $\cos(\theta)$ (top), in
signal (middle), and in distance to core (bottom). In the three cases,
it is almost constant and close to unity, which shows that our
model for the time uncertainty is in good agreement with the
experimental data. 
It is important to remark
that our model does not depend directly on the distance
of the tank to the shower core. Therefore, the results obtained in
this case can be regarded as an additional proof of validity.

\begin{figure}[t]
  \vspace{-2.5cm}
      \begin{center}
	\includegraphics*[width=0.7\textwidth]{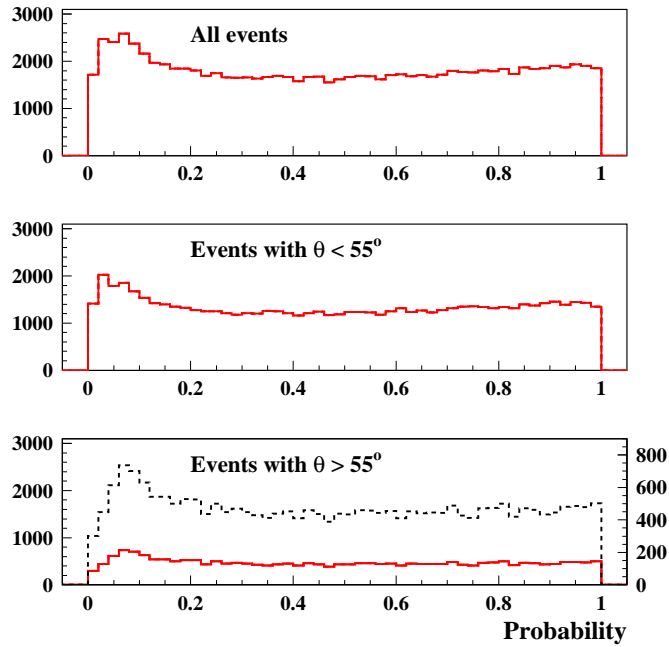}
	\vspace{-0.5cm}
	\caption {The $\chi^2$ probability distribution for all events
	  (top), events with zenith angle smaller than 55$^\circ$
	  (middle), and events with zenith angle larger than 55$^\circ$
	  (bottom). In the last figure the distribution is plotted with two
	  different scales, the same than the others (full line)
	  for comparison reasons and a zoom (dashed line) to see
	  the details. \label{fig:proba}}
      \end{center}
\end{figure}

We also studied the distribution of the $\chi^2$ probability of the T5
events~\cite{trigger} with 4 or more stations. In
figure~\ref{fig:proba} we show this distribution for all events (top),
for events with zenith angle 
smaller than 55$^\circ$ (middle) and events with zenith angle larger
than 55$^\circ$ (bottom)\footnote{We only plot probabilities larger
than 1\% to avoid the large peak at zero corresponding to badly
reconstructed events, which corresponds to about 9\% of the total
data.}. For the three cases, the distributions are
almost flat as it should be in the ideal case. It is important that
the flatness is observed both for large and small zenith angles separately,
which means that the model for the time uncertainty works well for all 
angles without compensating one set from the other. The flatness of these
distributions indicates that our model properly reproduces the
experimental uncertainty. It also indicates that the shower front model 
used in our fit (parabolic front with adjusted radius) is adequate.

\begin{figure}[t] 
       \begin{center}
	 \vspace{0.2cm}
	\includegraphics*[width=0.65\textwidth]{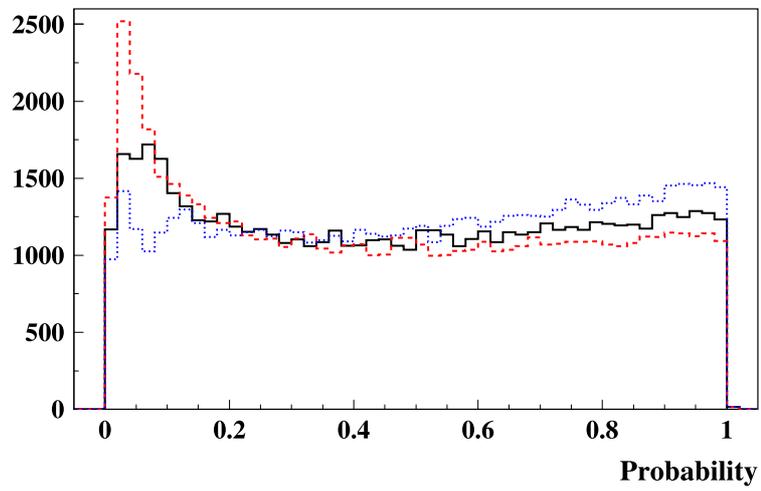}
	\caption{\small {Distribution of the probability of $\chi^2$, in
	    black (solid) the original one, in red (dashed) 1.2~$\chi^2$ and in
	    blue (dot) 0.8~$\chi^2$.}
	  \label{fig:proba20}}
      \end{center}
\end{figure}

Finally, to illustrate the sensitivity of the $\chi^2$ probability
distribution to the value of the variance used in the fit, we plotted
in figure~\ref{fig:proba20} the distribution obtained when one 
artificially varies the overall variance (e.g. $\chi^2$) normalization. 
In black (solid), the original curve, in red (dashed)
for a variance reduced by 20\% (individual $\chi^2$ mutliplied by
1.2), in blue (dotted) increased by 20\% (individual $\chi^2$ multiplied by
0.8). This 20\% in $\chi^2$ is equivalent to a 10\% change in $\sigma$. We
therefore conclude that we properly reproduce the measurement
uncertainties within 10\% level.

\section{Conclusions}

We developed a model to describe the measurement uncertainties of the
arrival direction of the shower front for an air-shower surface
detector array. This model is used for the Auger WCD and is based on 
the shower zenith angle, integrated signal and rise time measured in
the tanks. Absolute predictions of our model parameters for the time 
uncertainty measurements are consistent with the ones obtained from
analysis of adjacent tank data.  

We performed numerous tests to validate both our estimation methods
and our modeled values. All cases indicated a correct procedure and 
a good agreement with the experimental data.

This model together with the proper description of the front geometry
will allow for an optimal determination of the shower arrival direction.
Properly taking into account our measurement uncertainties also allows 
for an event by event determination of the precision on the
reconstructed arrival direction. 

\section{Acknowledgments}

This work was supported by the Conselho Nacional de
Desenvolvimento Cient\'{\i}fico e Tecnol\'ogico (CNPq), Brazil,  and the
Centre National de la Recherche Scientifique, Institut National de
Physique Nucl\'eare et Physique des Particules (IN2P3/CNRS), France. 
LEO - International Research Group.

\newpage

\newpage

\newpage

\newpage

\newpage

\newpage

\end{document}